\documentclass{emulateapj}

\newcommand{\muHz}{\,\mu{\rm Hz}}
\newcommand{\rhomean}{\langle \rho_* \rangle}
\newcommand{\Msun}{{\rm M}_\odot}
\newcommand{\Lsun}{{\rm L}_\odot}
\newcommand{\Rsun}{{\rm R}_\odot}
\def\rbf{\rm}
\def\fig{.}
\def\lwig{{\leavevmode\kern0.3em\raise.3ex\hbox{$<$}
\kern-0.8em\lower.7ex \hbox{$\sim$}\kern0.3em}}

\shortauthors{Christensen-Dalsgaard et al.}

\begin{document}

\title{Asteroseismic Investigation of Known Planet Hosts in the {\it Kepler}
Field}

\author{J. Christensen-Dalsgaard\altaffilmark{1,2}}
\author{H. Kjeldsen\altaffilmark{1,2}}
\author{T. M. Brown\altaffilmark{3}}
\author{R. L. Gilliland\altaffilmark{4}}
\author{T. Arentoft\altaffilmark{1,2}}
\author{S. Frandsen\altaffilmark{1,2}}
\author{P.-O. Quirion\altaffilmark{1,2,5}}
\author{W. J. Borucki\altaffilmark{6}}
\author{D. Koch\altaffilmark{6}}
\author{J. M. Jenkins\altaffilmark{7}}

\altaffiltext{1}{Department of Physics and Astronomy, Aarhus University, 
DK-8000 Aarhus C, Denmark: e-mail jcd@phys.au.dk}
\altaffiltext{2}{Danish AsteroSeismology Centre}
\altaffiltext{3}{Las Cumbres Observatory Global Telescope, Goleta, CA 93117}
\altaffiltext{4}{Space Telescope Science Institute, 3700 San Martin Drive,
Baltimore, MD 21218}
\altaffiltext{5}{Canadian Space Agency, 6767 Route de l'A\'eroport,
Saint-Hubert, QC, J3Y 8Y9 Canada (present address)}
\altaffiltext{6}{NASA Ames Research Center, MS 244-30, Moffett Field,
 CA 94035, USA}
\altaffiltext{7}{SETI Institute/NASA Ames Research Center,
 MS 244-30, Moffett Field, CA 94035, USA}

\begin{abstract}
In addition to its great potential for characterizing extra-solar 
planetary systems
the {\it Kepler mission} is providing unique data on stellar oscillations.
A key aspect of {\it Kepler} asteroseismology is the application
to solar-like oscillations of main-sequence stars.
As an example we here consider an initial analysis
of data for three stars in the {\it Kepler} field for 
which planetary transits were known from ground-based observations.
For one of these, HAT-P-7, we obtain a detailed frequency spectrum
and hence strong constraints on the stellar properties.
The remaining two stars show definite evidence for solar-like oscillations,
yielding a preliminary estimate of their mean densities.
\end{abstract}


\keywords{stars: fundamental parameters --- stars: oscillations --- 
planetary systems}


\section{Introduction}

%
The main goal of the {\it Kepler mission} is to characterize extra-solar
planetary systems, particularly Earth-like planets in the habitable zone
\citep[e.g.,][]{Boruck2009}.
The mission detects the presence of planets through the minute reduction
of the light from a star as a planet crosses the line of sight.
Several observations of such reductions at fixed time intervals
for a given star, and extensive follow-up observations, are used to
verify that the effect results from planet transits and to characterize 
the planet.
To ensure a reasonable chance of detection {\it Kepler} observes more
than 100,000 stars simultaneously, in a fixed field in the Cygnus-Lyra region.
Most stars are observed at a cadence of 29.4 min, but a subset
of up to 512 stars can be observed at a short cadence (SC) of 58.85\,s.
{\it Kepler} was launched on 6 March 2009 and data from the commissioning 
period and the first month of regular observations are now available.

The very high photometric accuracy required to detect planet transits
\citep{Boruck2010, Koch2010} also makes the {\it Kepler} observations
of great interest for asteroseismic studies of stellar interiors.
In particular, the SC data allow investigations of solar-like oscillations
in main-sequence stars.
Apart from the great astrophysical interest of such investigations
they also provide powerful tools to characterize stars that host
planetary systems \citep{Kjelds2009}.

In stars with effective temperature $T_{\rm eff} \lwig 7000\, {\rm K}$
we expect to see oscillations similar to those observed in the Sun
\citep[e.g.,][]{Christ2002}, excited stochastically by the near-surface
convection.
These are acoustic modes of high radial order;
in main-sequence stars such modes approximately satisfy the asymptotic relation
\begin{equation}
\nu_{nl} \simeq \Delta \nu_0 (n + l/2 + \epsilon) -l(l+1) D_0 \; 
\label{eq:asymp}
\end{equation}
\citep{Vandak1967, Tassou1980}.
Here $\nu_{nl}$ is the cyclic frequency,
$n$ is the radial order of the mode and $l$ is the degree, $l = 0$
corresponding to radial (i.e., spherically symmetric) oscillations.
Also, $\Delta \nu_0$ is essentially the inverse sound travel time across
the stellar diameter;
this is closely related to the mean stellar density
$\rhomean$: $\Delta \nu_0 \propto \rhomean^{1/2}$.
$D_0$ depends sensitively on conditions near the
center of the star; for stars during the central hydrogen burning phase
this provides a measure of stellar age. 
Finally, $\epsilon$ is determined by conditions near the stellar surface.
This regular form of the frequency spectrum simplifies 
the analysis of the observations, and the close relation between the
stellar properties and the parameters characterizing the frequencies make
them efficient diagnostics of the properties of the star.
{\rbf This has been demonstrated in the last few years through observations of
solar-like oscillations from the ground and from space
\citep[for reviews, see][]{Beddin2008, Aerts2009, Gillil2010a}.}

\begin{table*}
\begin{center}
\caption{Properties of transiting systems.\label{tbl-0}}
\begin{tabular}{cccccccccccc}
\tableline\tableline
Name & KIC No & $T_{\rm eff}$ (K) & [Fe/H] & $L/\Lsun$ & $\log(g)$ (cgs) &
$v \sin i$ & Source \\
 & & & & & & $({\rm km\,s^{-1}})$ & \\
\tableline
HAT-P-7 & 10666592 & $6350 \pm 80$ & $0.26 \pm 0.08$ & 
$4.9 \pm 1.1$ & $4.07 \pm 0.06$ & $3.8 \pm 0.5$ & (a) \\
        &              & $6525 \pm 61$ & $0.31 \pm 0.07$ &
                      & $4.09 \pm 0.08$ & & (b) \\
HAT-P-11 & 10748390 & $4780 \pm 50$ & $0.31 \pm 0.05$ &
$0.26 \pm 0.02$ & $4.59 \pm 0.03$ & $1.5 \pm 1.5$ & (c) \\
TrES-2 & 11446443 & $5850 \pm 50$ & $-0.15 \pm 0.10$ &
$1.17 \pm 0.10$ & $4.4 \pm 0.1$ & $2 \pm 1$ & (d) \\
       &              & $5795 \pm 73$ & $0.06 \pm 0.08$ &
                & $4.30 \pm 0.13$ & & (b) \\
\tableline
\end{tabular}
\tablecomments{Sources: (a): \citet{Pal2008}; (b): \citet{Ammler2009};
(c): \citet{Bakos2010}; (d): \citet{Sozzet2007}. In some cases 
asymmetric error bars have been symmetrized.
}
\end{center}
\end{table*}

Even observations allowing a determination of $\Delta\nu_0$ provide
useful constraints on $\rhomean$.
With a reliable determination of individual frequencies
$\rhomean$ is tightly constrained and an estimate of the stellar age can
be obtained.
This can greatly aid the interpretation of observations of planetary
transits \citep[e.g.,][]{Gillil2010b, Nutzma2010}.
We note that photometric observations such as those carried out by
{\it Kepler} are predominantly sensitive to modes of degree $l = 0{-}2$.
As indicated by Eq.~(\ref{eq:asymp}) these are sufficient to obtain information
about the core properties of the star.


Ground-based transit observations have identified three planetary systems
in the {\it Kepler} field: TrES-2 \citep{ODonov2006, Sozzet2007},
HAT-P-7 \citep{Pal2008}, and HAT-P-11 \citep{Dittma2009, Bakos2010}.
These systems have been observed by {\it Kepler} in SC mode.
Their properties (cf.\ Table~\ref{tbl-0})
indicate that they should display solar-like oscillations at 
observable amplitudes,
and hence they are obvious targets for {\it Kepler} asteroseismology.
Here we report the results of a preliminary asteroseismic
characterization of 
the central stars in the systems, based on the early {\it Kepler} data.


\section{Observations and data analysis}

\label{sec:obs}
%
We have analyzed data from {\it Kepler} for three planet-hosting stars
using a pipeline developed for fast and robust
analysis of all {\it Kepler} p-mode data \citep{CDetal2008, Huber2009}.
Each time series contains 63324 data points. 
SC data characteristics and minor post-pipeline processing are
discussed in \citet{Gillil2010c}. 
In addition a limb-darkened transit
light curve model fit has been removed and 5-$\sigma$ clipping applied
to remove outlying data points from each of the time series.
The frequency analysis contains four main steps:

\begin{figure}
\epsscale{1.0}
\plotone{\fig/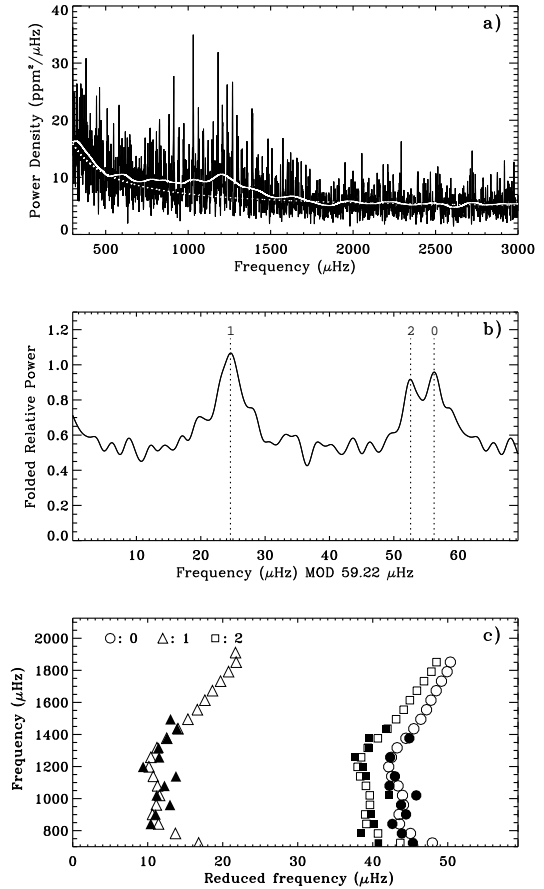}
\caption{
(a) Power spectrum of HAT-P-7 for frequencies between 300 and $3000 \muHz$. The
spectrum is smoothed with a gaussian filter with a FWHM of $3 \muHz$. The noise
level at high frequencies corresponds to 1.1 ppm in amplitude. The white curve
is a smoothed power spectrum with a gaussian filter ($150 \muHz$ FWHM). A fit
to the background (dashed white curve) is also shown. The excess power and the
individual p-modes are evident.
(b) Folded power spectrum, {\rbf between 750 and $1500 \muHz$},
for HAT-P-7 for a large separation of $59.22 \muHz$.
Indicated are the positions corresponding
to radial modes ($l=0$) and non-radial modes with $l=1$ and 2.
The measured positions are used to identify the individual oscillation
modes in panel (a).
(c) \'Echelle diagram (see text) for frequencies of
degree $l = 0$, 1, and 2 in HAT-P-7;
a frequency separation of $59.36 \muHz$ and a starting frequency of
$10.8 \muHz$ were used.
The filled symbols, coded for degree as indicated, show the observed
frequencies, while the open symbols are for Model 3 in Table~\ref{tbl-2},
minimizing $\chi_\nu^2$.
\label{fig:obs}}
\end{figure}

\begin{enumerate}

\item We calculate an oversampled (factor of four) version of the
power spectrum by using a least-squares fitting. We smoothed the
spectrum to $3 \muHz$ resolution to remove the fine structure
caused by the finite mode lifetime.

\item We correlated the smoothed power spectrum
with an equally spaced comb of delta functions,
separated by $\Delta \nu_0/2$, and confined to a Gaussian-shaped band 
with a full width at half maximum of $5 \Delta \nu_0$.
We adopted the maximum of this
convolution over lags between 0 and 0.5 $\Delta \nu_0$ as the filter output
for each $\Delta \nu_0$.

\item After identifying the peak correlation for the best matched model
filter and extracting the large separation corresponding to this
peak we calculate the {\it folded spectrum} (see Fig.~\ref{fig:obs}b), i.e.,
the sum of the power as a function of frequency
modulo the optimum large separation (the one corresponding to the
peak correlation). The summed power
is used to locate the p-mode structure and identify the ridges
corresponding to the different mode degrees
(based on the asymptotic relation).

\item From the asymptotic relation and the identification of
mode degrees we finally identify the position of the individual
p-mode frequencies in the smoothed version of the power spectrum;
when more than one mode is seen near the expected frequency
we use the power-weighted average of the two peaks.
Those extracted frequencies and the mode identifications are used
in the modeling.

\end{enumerate}

For observations with low signal-to-noise ratio it may not be possible
to identify the individual frequencies. 
In such cases the analysis is carried through step 2, to determine the maximum
response and hence an estimate of the large separation.

Results on the three individual cases are presented in \S\ref{sec:results}.

\section{Model fitting}

\label{sec:model}
%
Stellar evolution models and adiabatic oscillation frequencies were
computed using the Aarhus codes \citep{Christ2008a, Christ2008b},
with the OPAL equation of state \citep{Rogers1996} and opacity
\citep{Iglesi1996} and the NACRE nuclear reaction parameters
\citep{Angulo1999}.
In some cases (see below) diffusion and settling of helium were included,
using the simplified formulation of \citet{Michau1993}.
Convection was treated with the \citet{Bohm1958} mixing-length formulation,
with a mixing length $\alpha_{\rm ML} = 2.00$ in units of the pressure
scale height roughly corresponding to a solar calibration.
In some models with convective cores, overshoot was included over
a distance of $\alpha_{\rm ov}$ pressure scale heights.
Evolution started from chemically homogeneous zero-age models.
The initial abundances by mass $X_0$ and $Z_0$ of hydrogen and
heavy elements were characterized by the assumed value of [Fe/H],
using as reference a present solar surface composition with
$Z_{\rm s}/X_{\rm s} = 0.0245$ \citep{Greves1993} and assuming,
from galactic chemical evolution, that $X_0 = 0.7679 - 3 Z_0$.

From the observed $\Delta \nu_0$, effective temperature and composition
an initial estimate of the stellar parameters was obtained using the 
grid-based SEEK pipeline (Quirion et al., in preparation).
Smaller grids were then computed in the vicinity of these initial parameters,
to obtain tighter constraints on stellar properties.
For HAT-P-7 the analysis of the observations yielded frequencies of
individually identified modes;
here the analysis was based on
\begin{equation}
\chi_\nu^2 = {1 \over N - 1}
\sum_{nl} \left( {\nu_{nl}^{\rm (obs)} - \nu_{nl}^{\rm (mod)}
\over \sigma_\nu} \right)^2 \; ,
\label{eq:chisqnu}
\end{equation}
where $\nu_{nl}^{\rm (obs)}$ and $\nu_{nl}^{\rm (mod)}$ are the observed
and model frequencies, $\sigma_\nu$ is the standard error in the observed
frequencies (assumed to be constant) and $N$ is the number of
observed frequencies.
In addition, we considered $\chi^2 = \chi_\nu^2 + \chi_T^2$, where
$\chi_T^2$ is the corresponding normalized square difference between the
observed and model effective temperature.
When $\chi_\nu^2$ was available we minimized it along each evolution track
and considered the resulting minimum values, and the corresponding value
of $\chi^2$, as a function of the parameters characterizing the models
\citep[see][for details]{Gillil2010b}.
When only the large separation $\Delta \nu_0$ could be determined from
the observations, we identified the model along each track which matched
$\Delta \nu_0$ and considered the resulting $\chi_T^2$ as a function
of the model parameters.

\section{Results}

\label{sec:results}
\subsection{HAT-P-7}

%
The observed power spectrum for HAT-P-7 is shown in Fig.~\ref{fig:obs}a.
The presence of solar-like p-mode peaks, with a maximum power around 
1.1\,mHz, is evident.
At high frequency the noise level in the amplitude spectrum is 
1.1 parts per million (ppm),
with some increase at lower frequency, likely due to the effects of stellar
granulation.

Carrying out the correlation analysis described in \S\ref{sec:obs}
we determined the large separation as $\Delta \nu_0 = 59.22 \muHz$.
Figure~\ref{fig:obs}b shows the resulting folded spectrum.
This clearly shows two closely spaced peaks, identified as corresponding
to modes of degree $l = 0$ and 2, and single peak separated from these
two by approximately $\Delta \nu_0/2$, corresponding to $l = 1$.
On this basis we finally determined the individual frequencies,
identifying the modes from the asymptotic relation;
the final set includes 33 p-mode frequencies, determined with a 
standard error $\sigma_\nu = 1.4 \muHz$.
These frequencies, {\rbf corresponding to radial orders between 11 and 24},
are illustrated in Fig.~\ref{fig:obs}c in an \'echelle diagram (see below).

\begin{table*}
\begin{center}
\caption{Stellar evolution models fitting the observed
frequencies for HAT-P-7.
\label{tbl-2}}
\begin{tabular}{cccccccccccc}
\tableline\tableline
$No$ & $M_*/{\rm M}_\odot$ & Age & $Z_0$ & $X_0$ & $\alpha_{\rm ov}$ & $R_*/{\rm R}_\odot$ &
$\rhomean$ & $T_{\rm eff}$ & $L_*/{\rm L}_\odot$ & $\chi_\nu^2$ & $\chi^2$ \\
     &                     & (Gyr)&      &       &          &          &
$({\rm g\,cm^{-3}})$ & (K) &  &  &   \\
\tableline
1 & 1.53 & 1.758 & 0.0270 & 0.6870 & 0.0 & 1.994 & 0.2718 & 6379 & 5.91 & 1.08 & 1.21 \\
2 & 1.52 & 1.875 & 0.0290 & 0.6809 & 0.1 & 1.992 & 0.2708 & 6355 & 5.81 & 1.04 & 1.04 \\
3 & 1.50 & 2.009 & 0.0270 & 0.6870 & 0.2 & 1.981 & 0.2718 & 6389 & 5.87 & 1.00 & 1.24 \\
\tableline
\end{tabular}
\tablecomments{Models minimizing $\chi_\nu^2$ (cf.\ Eq.~\ref{eq:chisqnu})
along the evolution tracks,
illustrated in Fig.~\ref{fig:HR}.
The models have been selected as providing the smallest $\chi_\nu^2$ for
each of the three values of the overshoot parameter $\alpha_{\rm ov}$.
The smallest value of $\chi_\nu^2$ is obtained for Model 3.}
\end{center}
\end{table*}

A grid of models was computed for masses between $1.41$ and $1.61 \,\Msun$,
[Fe/H] between 0.17 and 0.38, and $\alpha_{\rm ov} = 0, 0.1$ and $0.2$,
extending well beyond the end of central hydrogen burning.
The modeling did not include diffusion and settling.
At the mass of this star the outer convection zone is
quite thin, and as a result the settling timescale is much shorter than
the age of the star.
Including settling, without compensating effects such as partial mixing in
the radiative region or mass loss, leads to a rapid change in the surface
composition which is inconsistent with the observed [Fe/H];
for simplicity we therefore neglected these effects for HAT-P-7.%
\footnote{Artificially suppressing settling in the outer layers,
while including diffusion and settling in the core, leads to results that
are very similar to those presented here.}

\begin{figure}
\epsscale{1.0}
\plotone{\fig/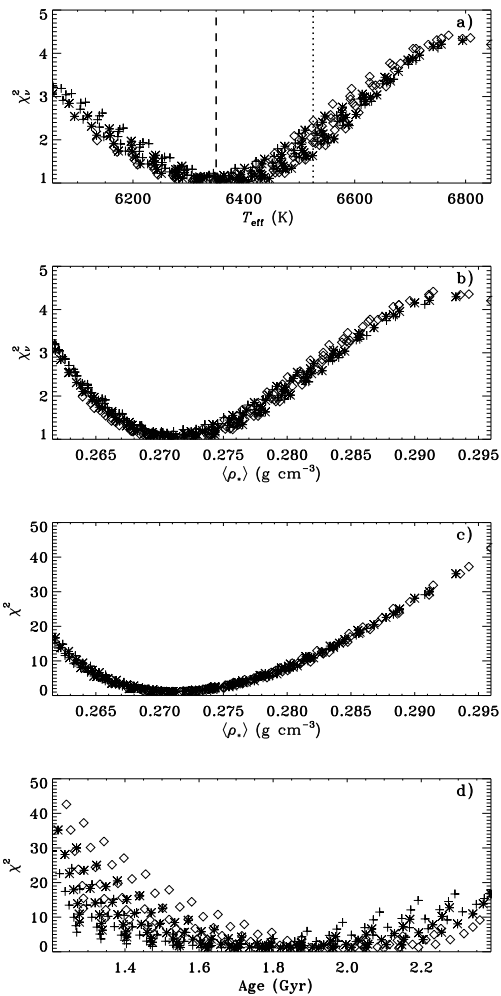}
\caption{
Results of fitting the observed frequencies
to a grid of stellar models (see text for details).
Plusses, stars and diamonds correspond to models with $\alpha_{\rm ov} = 0$
(no overshoot), 0.1, and 0.2.
(a) Minimum mean square deviation $\chi_\nu^2$
of the frequencies (cf.\ Eq.~\ref{eq:chisqnu}) along each evolution track,
against the effective temperature $T_{\rm eff}$ of the corresponding models.
The vertical dashed and dotted lines indicate the effective temperatures
found by \citet{Pal2008} and \citet{Ammler2009}.
(b) Minimum mean square deviation $\chi_\nu^2$
against the mean density $\rhomean$ of the corresponding models.
{\rbf (c) As (b), but showing the combined $\chi^2$.
(d) $\chi^2$ against the age for the models that
minimize $\chi_\nu^2$;}
the different ridges correspond to the different masses in the grid,
the more massive models resulting in a lower estimate of the age.
\label{fig:chisq}}
\end{figure}

The computed frequencies were corrected according to the procedure of
\citet{Kjelds2008} for errors in the modeling of the near-surface layers,
by adding $a (\nu/\nu_0)^b$ where $a = 0.1158 \muHz$,
$\nu_0 = 1000 \muHz$ and $b = 4.9$. 
As discussed in \S\ref{sec:model}, for each evolution track, 
characterized by a set of model parameters, we minimized
the departure $\chi_\nu^2$ of the model frequencies
from the observations, defining the best model for this set.

We first consider $\chi_\nu^2$ as a function of the effective temperature of
the models (Fig.~\ref{fig:chisq}a).
It is evident that there is a clear minimum in $\chi_\nu^2$;
this is consistent with the determination of $T_{\rm eff}$ by
\citet{Pal2008} but not with the somewhat higher temperature
obtained by \citet{Ammler2009} (see also Table~\ref{tbl-0}).
Thus in the following we use the observed quantities from \citet{Pal2008}.

Since the frequencies to leading order are determined by the mean
stellar density $\rhomean$,
Fig.~\ref{fig:chisq}b,c show $\chi_\nu^2$ and $\chi^2$ as functions
of $\rhomean$.
It is evident that the best-fitting models occupy a narrow range 
of $\rhomean$, with a well-defined minimum.
Fitting a parabola to $\chi^2$ in panel (c) we 
obtain the estimate $\rhomean = 0.2712 \pm 0.0032 \,{\rm g \, cm^{-1}}$.
In Fig.~\ref{fig:chisq}d $\chi^2$ is shown against model age.
Here the variation with model parameters is substantially stronger,
resulting in a greater spread in the inferred age;
in particular, it is evident, not surprisingly, that the results
depend on the extent of convective overshoot.
From the figure we estimate that the age of HAT-P-7 is between
1.4 and 2.3\,Gyr.

\begin{figure}
\epsscale{1.0}
\plotone{\fig/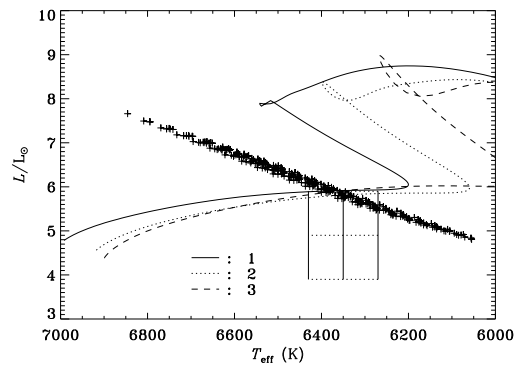}
\caption{Theoretical HR diagram with selected evolutionary tracks,
corresponding to the models defined in Table 2. 
The '+' indicate the models
along the full set of evolutionary sequences minimizing the
difference between the computed and observed frequencies.
The box is centered on the
$L$ and $T_{\rm eff}$ as given by \citet{Pal2008},
with a size matching the errors on these quantities.
\label{fig:HR}}
\end{figure}


Examples of evolution tracks are shown in Fig.~\ref{fig:HR};
parameters for these models are provided in Table~\ref{tbl-2}. 
They were chosen to give the smallest $\chi_\nu^2$ for each
of the three values of $\alpha_{\rm ov}$ considered.
Also shown are the locations of the models minimizing $\chi_\nu^2$ along
each of the computed tracks; 
these evidently fall close to a line in the HR diagram, corresponding to
the small range in $\rhomean$.
The range of luminosities, from \citet{Pal2008}, is based on modeling and
hence has not been used in our fit; even so, it is gratifying that the
present models are essentially consistent with these values.
Also, as indicated by Fig.~\ref{fig:chisq}a and Table~\ref{tbl-2},
the best-fitting models are close to the value of $T_{\rm eff}$
obtained by \citet{Pal2008}.

The match of the best-fitting model (Model 3 of Table~\ref{tbl-2}) to the
observed frequencies is illustrated in a so-called {\it \'echelle diagram}
\citep{Grec1983} in Fig.~\ref{fig:obs}c.
In accordance with Eq.~(\ref{eq:asymp}) the frequency spectrum is divided into
slices of length $\Delta \nu$, starting at a frequency of $10.8 \muHz$;
the figure shows the location of the observed (filled symbols) and computed
(open symbols) frequencies within each slice,
against the starting frequency of the slice; 
{\rbf the model results extend to the
acoustical cut-off frequency, $1930 \muHz$, of the model.}
There is clearly a very good overall agreement between model and observations,
including the detailed variation with frequency which reflects the frequency
dependence of the large separation, as a possible diagnostics of the outer
layers of the star \citep[e.g.,][]{Houdek2007}.

We have finally made a fit of the inferred $\rhomean$, as well as 
$T_{\rm eff}$ and [Fe/H] from \citet{Pal2008}, to computed evolutionary
tracks from the Yonsei-Yale compilation 
\citep{Yi2001}.
This was based on a Markov Chain Monte Carlo analysis to obtain the
statistical properties of the inferred quantities
\citep[see][for details]{Brown2010}.
This resulted in $M = 1.520 \pm 0.036 \,\Msun$, $R = 1.991 \pm 0.018 \,\Rsun$
and an age of $2.14 \pm 0.26$\,Gyr.
We note that the age estimate reflects the specific assumptions in the
Yonsei-Yale evolution calculations; 
as indicated by Fig.~\ref{fig:chisq}d the true uncertainty in 
the age determination is likely somewhat larger.

\subsection{HAT-P-11}

%
For HAT-P-11 the oscillation amplitudes were much smaller than in HAT-P-7,
as expected from the general scaling of amplitudes with stellar mass and
luminosity \citep[e.g.,][]{Kjelds1995}.
Thus with the present short run of data it has only been possible to determine
the large separation $\Delta \nu_0 = 180.1 \muHz$ from the maximum in
the correlation analysis.
We have matched this to a grid of models, including diffusion and
settling of helium, with masses between 
0.7 and $0.9 \, \Msun$ and [Fe/H] between 0.21 and 0.41.
These models provide a good fit to the observed $T_{\rm eff}$ and $L/\Lsun$;
note that in the present case the luminosity is based on a reasonably
well-determined parallax.
We have determined an estimate of $\rhomean$ by averaging the results
of those models which match the observed $\Delta \nu_0$
and lie within 2 standard deviations ($\pm 100$\,K)
from the value of $T_{\rm eff}$ provided by \citet{Bakos2010};
the result is $\rhomean = 2.5127 \pm 0.0009 \, {\rm g \, cm^{-3}}$.
Although the formal error is extremely small, owing to a tight relation
between the large separation and the mean density for stars in this region
in the HR diagram, the true error is undoubtedly substantially larger.
In particular, we neglected the error in the determination
of $\Delta \nu_0$ and these data have not allowed a correction for
the systematic errors in the modeling of the near-surface layers of the star.

\subsection{TrES-2}

Here also we were unable to determine individual frequencies from the
present set of data.
The expected amplitudes are smaller than for HAT-P-7, and the noise
level higher due to the fainter magnitude of TrES-2.
The correlation analysis yielded two possible values of $\Delta \nu_0$:
$97.7 \muHz$ and $130.7 \muHz$.
For this star $\rhomean$ has been determined from the analysis of the
transit light curve.
\citet{Sozzet2007} obtained $\rhomean = 1.375 \pm 0.065 \, {\rm g \, cm^{-3}}$,
while \citet{Southw2009} found $\rhomean = 1.42 \pm 0.13 \,{\rm g \, cm^{-3}}$.
From the scaling with $\rhomean^{1/2}$ the smaller of the two possible values
of $\Delta \nu_0$ is clearly inconsistent with these values of $\rhomean$,
while $\Delta \nu_0 = 130.7 \muHz$ yields models that are consistent with
the observed $T_{\rm eff}$ and $\log(g)$ of \citet{Sozzet2007}
as well as with these values of the mean density.
Here we considered a grid of models with helium diffusion and settling,
masses between 0.85 and $1.1 \, \Msun$ and [Fe/H] between $-0.25$ and $-0.05$.
Determining again the mean value of $\rhomean$ for those models that matched
$\Delta \nu_0$ and had $T_{\rm eff}$ within two standard deviations of
the value of \citet{Sozzet2007} we obtained
$\rhomean = 1.3233 \pm 0.0027 \, {\rm g \, cm^{-3}}$.
As in the case of HAT-P-11 the true error is likely substantially higher.

\section{Discussion and conclusion}

%
The present preliminary analysis provides a striking demonstration of the
potential of {\it Kepler} asteroseismology and its supporting role
in the analysis of planet hosts.
These stars will undoubtedly be observed throughout the mission and
hence the quality of the data will increase substantially.
For HAT-P-7 the detected frequencies are already close to what will
be required for a detailed analysis of the stellar interior, beyond the
determination of the basic parameters of the star.
Thus here we can look forward to a test of the assumptions of the
stellar modeling;
the resulting improvements will further constrain the overall properties
of the star, in particular its age.
Also, given the observed $v \sin i$ we expect a rotational splitting comparable
to that observed in the Sun, and hence likely detectable with a few months
of observations.
For the other two stars there is strong evidence for the presence of solar-like
oscillations;
thus continued observations will very likely result in the determination of
individual frequencies and hence further constraints on the properties of
the stars.

\acknowledgements
Funding for this Discovery mission is provided by NASA's Science Mission
Directorate.
We are very grateful to the entire {\it Kepler} team, whose efforts have
led to this exceptional mission.
The present work was supported by the Danish Natural Science Research Council.

{\it Facilities:} \facility{The Kepler Mission}




\end{document}